\documentclass[aps,prl,twocolumn,preprintnumbers,nofootinbib,longbibliography,floatfix]{revtex4-1}
\usepackage{multirow,graphicx,color,float,hyperref,amsmath,amssymb}

\definecolor{darkblue}{rgb}{0,0,0.7}
\definecolor{darkred}{rgb}{0.7,0,0}
\definecolor{darkgreen}{rgb}{0,0.7,0}
\definecolor{orange}{rgb}{1, 0.08, 0.6}
\definecolor{purple}{rgb}{0.4, 0.04, 0.2}

\definecolor{jdtcolor}{rgb}{0.8392,0.1529,0.1569}
\definecolor{emmcolor}{rgb}{0.1725,0.6275,0.1725}
\definecolor{ptkcolor}{rgb}{0.1016,0.4466,0.6859}

\DeclareRobustCommand{\Fig}[1]{Fig.~\ref{#1}}

\DeclareRobustCommand{\Eq}[1]{Eq.~(\ref{#1})}

\DeclareRobustCommand{\Ref}[1]{Ref.~\cite{#1}}
\DeclareRobustCommand{\Refs}[1]{Refs.~\cite{#1}}

\begin{document}
\title{The Metric Space of Collider Events}

\author{Patrick T. Komiske}
\email{pkomiske@mit.edu}

\author{Eric M. Metodiev}
\email{metodiev@mit.edu}

\author{Jesse Thaler}
\email{jthaler@mit.edu}

\affiliation{Center for Theoretical Physics, Massachusetts Institute of Technology, Cambridge, MA 02139, USA}
\affiliation{Department of Physics, Harvard University, Cambridge, MA 02138, USA}

\begin{abstract}
When are two collider events similar?
Despite the simplicity and generality of this question, there is no established notion of the distance between two events.
To address this question, we develop a metric for the space of collider events based on the earth mover's distance: the ``work'' required to rearrange the radiation pattern of one event into another.
We expose interesting connections between this metric and the structure of infrared- and collinear-safe observables, providing a novel technique to quantify event modifications due to hadronization, pileup, and detector effects.
We showcase how this metrization unlocks powerful new tools for analyzing and visualizing collider data without relying upon a choice of observables.
More broadly, this framework paves the way for data-driven collider phenomenology without specialized observables or machine learning models.
\end{abstract}

\preprint{MIT-CTP 5102}

\maketitle

High-energy particle collisions produce a tremendous number of intricately correlated particles, especially when energetic quarks and gluons are involved.
Behind this apparent complexity, however, the overall flow of energy in an event is a robust memory of its simpler partonic origins~\cite{Sterman:1977wj,Georgi:1977sf,Donoghue:1979vi,Altarelli:1981ax,Dokshitzer:1991eq,Tkachov:1995kk,Sveshnikov:1995vi,Hofman:2008ar}.
Surprisingly, no definition of the similarity between events presently exists that sharply captures this correspondence.
In the absence of a metric, efforts typically fall back upon ad hoc methods such as comparing specific observables~\cite{Cacciari:2007fd,Cacciari:2014gra,Bertolini:2014bba,Martinez:2018fwc,Komiske:2017ubm} or matching the pixels of calorimeter images~\cite{Komiske:2017ubm,Cogan:2014oua,deOliveira:2015xxd,Paganini:2017hrr,Paganini:2017dwg}.
These approaches suffer from significant pathologies: disparate event topologies can give rise to identical observable values, while pixels lack stability under small perturbations.
A theoretically and experimentally robust definition of the ``distance'' between events would profoundly expand our ability to explore the structure of collider data and unlock entirely new ways to probe events.

In this letter, we advocate for the earth (or energy) mover's distance (EMD)~\cite{DBLP:journals/pami/PelegWR89,Rubner:1998:MDA:938978.939133,Rubner:2000:EMD:365875.365881,DBLP:conf/eccv/PeleW08,DBLP:conf/gsi/PeleT13} as a metric for the space of collider events.
We propose a variant of the EMD, inspired by \Refs{DBLP:conf/eccv/PeleW08,DBLP:conf/gsi/PeleT13}, that allows events with different total energies to be sensibly compared.
The EMD is the minimum ``work'' required to rearrange one event $\mathcal E$ into the other $\mathcal E'$ by movements of energy $f_{ij}$ from particle $i$ in one event to particle $j$ in the other:
\vspace{-0.5em}
\begin{align}\label{eq:emd}
&\text{EMD}(\mathcal E, \mathcal E') = \min_{\{f_{ij}\}}\sum_{ij} f_{ij} \frac{\theta_{ij}}{R} + \left|\sum_i E_i - \sum_j E'_j\right|,\\
&f_{ij} \ge 0,\,\,\, \sum_{j} f_{ij} \le E_i,\,\,\, \sum_i f_{ij} \le E'_j,\,\,\,\sum_{ij}f_{ij}= E_\text{min},\nonumber
\end{align}
where $i$ and $j$ index particles in events $\mathcal E$ and $\mathcal E'$, respectively, $E_i$ is the particle energy, $\theta_{ij}$ is an angular distance between particles, and $E_\text{min}=\min(\sum_i E_i, \sum_j E_j')$ is the smaller of the two total energies.
$R$ is a parameter that controls the relative importance of the two terms.
While energies and angles are used here for clarity, we will use transverse momenta $p_{T}$ and rapidity-azimuth $(y,\phi)$ distances for our applications relevant for the Large Hadron Collider (LHC).

\begin{figure}[t]
\centering
\includegraphics[width=\columnwidth]{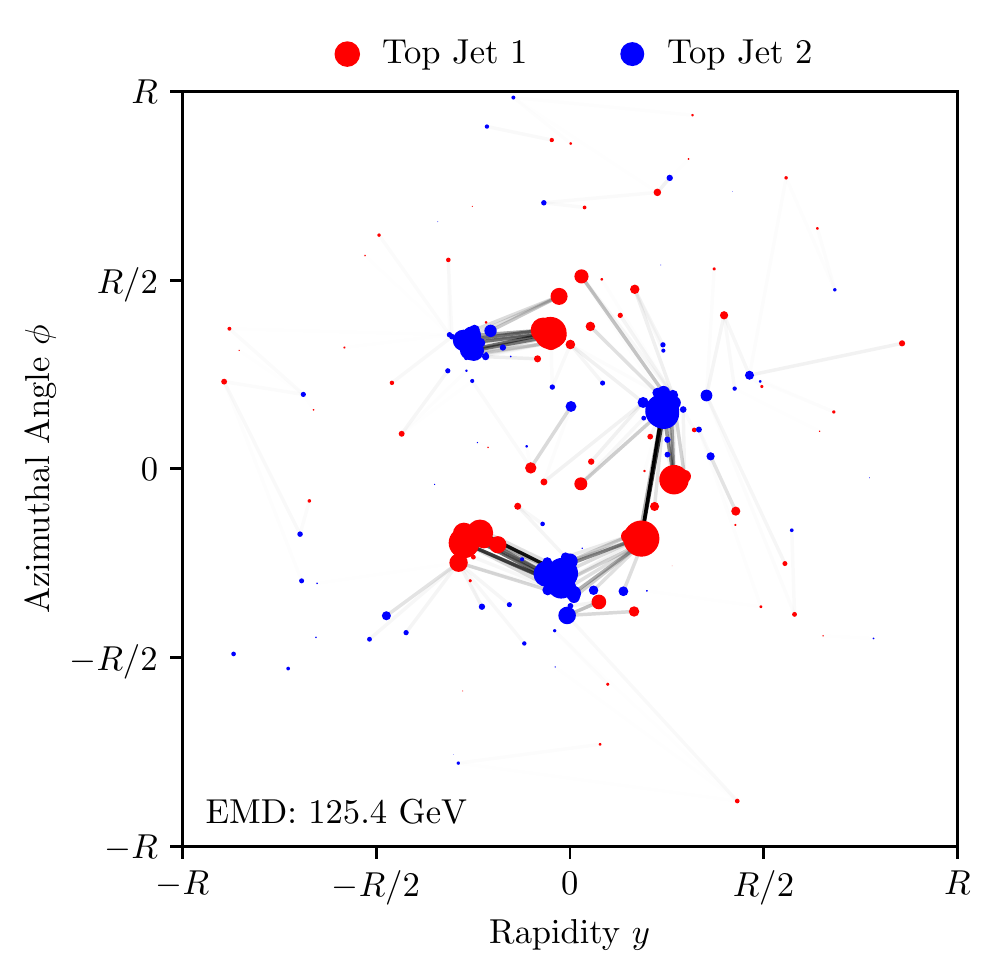}
\caption{The optimal movement to rearrange one top jet (red) into another (blue).
Particles are shown as points in the rapidity-azimuth plane with areas proportional to their transverse momenta.
Darker lines indicate more transverse momentum movement.
The energy mover's distance in \Eq{eq:emd} is the total ``work'' required to perform this rearrangement.}
\label{fig:exemd}
\end{figure}

The EMD that we propose in \Eq{eq:emd} has dimensions of energy, where the first term quantifies the difference between the two radiation patterns and the second term accounts for the creation or destruction of energy.
It is a true metric (satisfying the triangle inequality) as long as $\theta_{ij}$ is a metric and $R \ge \frac12 \theta_\text{max}$, where $\theta_\text{max}$ is the maximum attainable angular distance between particles.
For instance, $R$ must be at least the jet radius for conical jets.
Formally, the EMD metrizes the energy flow, as it treats events differing only by soft particles or collinear splittings identically.
This hints at a deep connection to infrared and collinear (IRC) safety of observables~\cite{Kinoshita:1962ur,Lee:1964is,Brock:1993sz,Weinberg:1995mt}, which we explore further below.

A metric for comparing events is particularly relevant for probing the substructure of jets~\cite{Seymour:1991cb,Seymour:1993mx,Butterworth:2002tt,Butterworth:2007ke,Butterworth:2008iy,Abdesselam:2010pt,Altheimer:2012mn,Altheimer:2013yza,Adams:2015hiv,Larkoski:2017jix,Asquith:2018igt}, collimated sprays of particles resulting from the fragmentation and hadronization of high-energy quarks and gluons via quantum chromodynamics (QCD).
Here, we will consider three classes of jets which have different intrinsic topologies: three-pronged boosted top quark jets, two-pronged boosted $W$ boson jets, and single-pronged QCD (quark or gluon) jets.
We generate proton-proton collision events at the LHC with \textsc{Pythia}~8.235~\cite{Sjostrand:2014zea} at $\sqrt{s}=14$ TeV including hadronization and multiple particle interactions.
Anti-$k_T$ jets~\cite{Cacciari:2008gp} with a jet radius of $1.0$ are clustered using \textsc{FastJet} 3.3.1~\cite{Cacciari:2011ma}, and up to two jets with $p_T\in[500,550]$ GeV and $|y|<1.7$ are kept.
This $p_T$ selection is representative of an intermediate energy range for jets at the LHC and allows for sensitivity to the effects of both terms in \Eq{eq:emd}.
Jets are longitudinally boosted and rotated to center the jet four-momentum at $(y,\phi) = 0$ as well as to vertically align the principal component of the constituent transverse momentum flow in the rapidity-azimuth plane; this removes the dependence of the EMD on these jet isometries.

We record the final-state hadrons, as well as the partons (before hadronization) and the hard $W$/top decay products, that are within a jet radius of the jet four-momentum.
We use the Python Optimal Transport~\cite{flamary2017pot} library to compute EMDs with the minimal choice of $R=1.0$, the jet radius.
The energy difference penalty in \Eq{eq:emd} is implemented using a fictitious particle at a distance $R$ from all other particles.
\Fig{fig:exemd} shows the optimal energy movement between two example top jets.

\begin{figure}[t]
\centering
\includegraphics[width=\columnwidth]{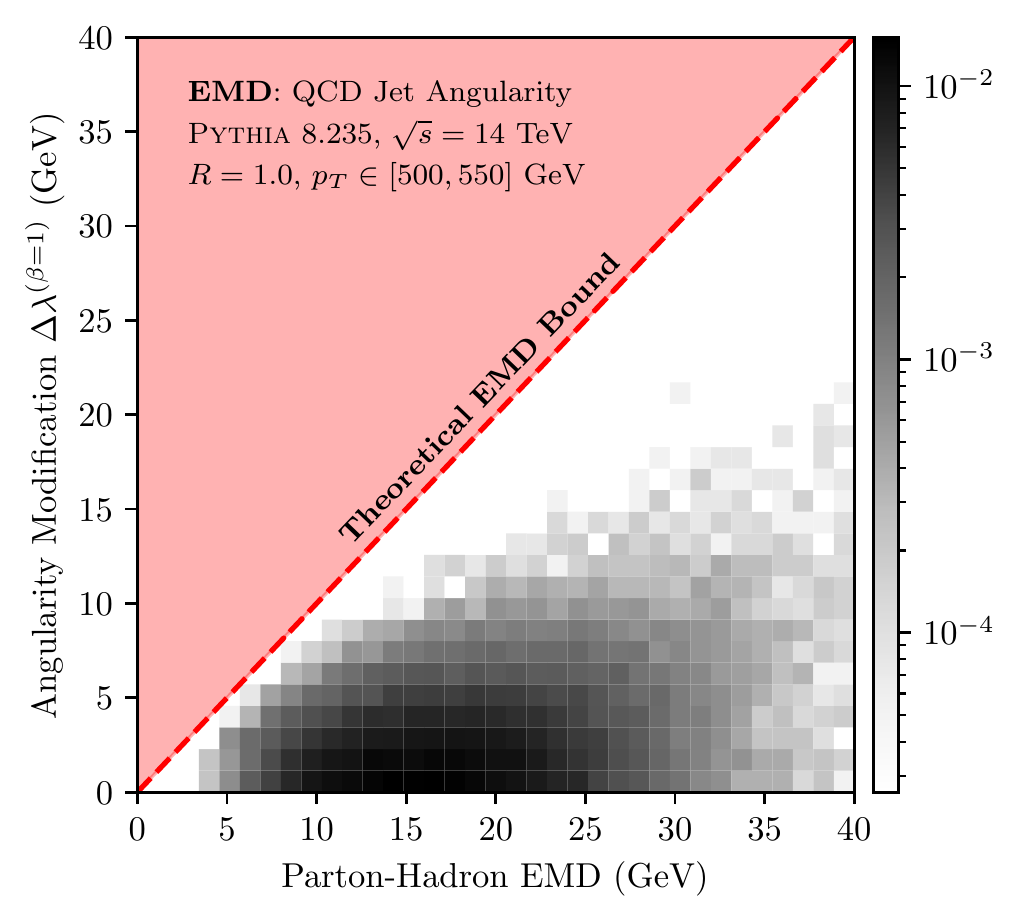}
\caption{Two-dimensional histogram of the EMD between 30k QCD jets before and after hadronization versus the corresponding $\beta=1$ angularity modification.
The red region is excluded based on the bound in \Eq{eq:dlam}, shown as a dashed red line.
The bound is clearly satisfied and is nearly saturated for EMD~$\lesssim10$~GeV.}
\label{fig:angemd}
\end{figure}

We begin by highlighting a remarkable mathematical property of the EMD which provides a quantitative understanding of an observable's sensitivity to the radiation pattern.
Specifically, we relate the EMD to additive IRC-safe observables via the Kantorovich-Rubinstein~\cite{kantorovich1958space} duality theorem.
Applying this theorem to our variant of the EMD, we derive the following mathematical bound between two events $\mathcal E$ and $\mathcal E'$:
\begin{equation}
\frac{1}{RL}\left|\sum_{i} E_i \Phi(\hat p_i) - \sum_{j} E_j' \Phi(\hat p_j')\right| \le \text{EMD}(\mathcal E, \mathcal E'),
\label{eq:krdual}
\end{equation}
where $i,\,j$ index $\mathcal E,\,\mathcal E'$, respectively, $\hat p_i$ is the particle angular position, and $\Phi$ is any $L$-Lipschitz function (essentially, with gradient size bounded by $L$) which vanishes at the center of the space (e.g.\ the jet axis).
The implications of \Eq{eq:krdual} are simple yet profound: the similarity of events according to the EMD metric guarantees the closeness of their $\mathcal O = \sum_{i=1}^M E_i \Phi(\hat p_i)$ observable values in a precise way that depends on $\Phi$.
By formulating IRC-safe observables in the language of additive energy-weighted structures~\cite{Komiske:2017aww,Komiske:2018cqr}, \Eq{eq:krdual} can be applied to provide a robust bound.

As a concrete example, we demonstrate how the EMD bounds hadronization modifications of jet angularities~\cite{Larkoski:2014pca} (see also \Refs{Berger:2003iw,Almeida:2008yp,Ellis:2010rwa,Larkoski:2014uqa}), $\lambda^{(\beta)} = \sum_i p_{T,i} \theta_i^{\beta}$ where $\theta_i$ is the rapidity-azimuth distance to the jet axis.
These angularities are evidently of the form in \Eq{eq:krdual} with $\Phi(y_i,\phi_i) = (y_i^2 + \phi_i^2)^{\beta/2}$, which for $\beta\ge1$ is a $\beta$-Lipschitz function over our $R=1.0$ jet cone, hence:
\begin{equation}
\label{eq:dlam}
\Delta\lambda^{(\beta)} = |\lambda^{(\beta)}(\mathcal E) -  \lambda^{(\beta)}(\mathcal E')| \le \beta\, \text{EMD}(\mathcal E, \mathcal E').
\end{equation}
The EMD between two events yields a robust upper bound of the difference in their $\beta\ge 1$ angularity values.
This bound is borne out in \Fig{fig:angemd}, where the angularity differences and EMDs are computed for the same QCD jets before and after hadronization.
For this jet $p_T$ range, hadronization modifies events by EMD~$\lesssim 30$~GeV and correspondingly modifies $\lambda^{(\beta=1)}$ by no more than this amount.
The intuitive picture of parton-hadron duality~\cite{Dokshitzer:1991eq}, that the energy flow in an event is robust to nonperturbative effects, is quantified by considering the EMD that these nonperturbative effects can induce.

\begin{figure}[t]
\centering
\includegraphics[width=\columnwidth]{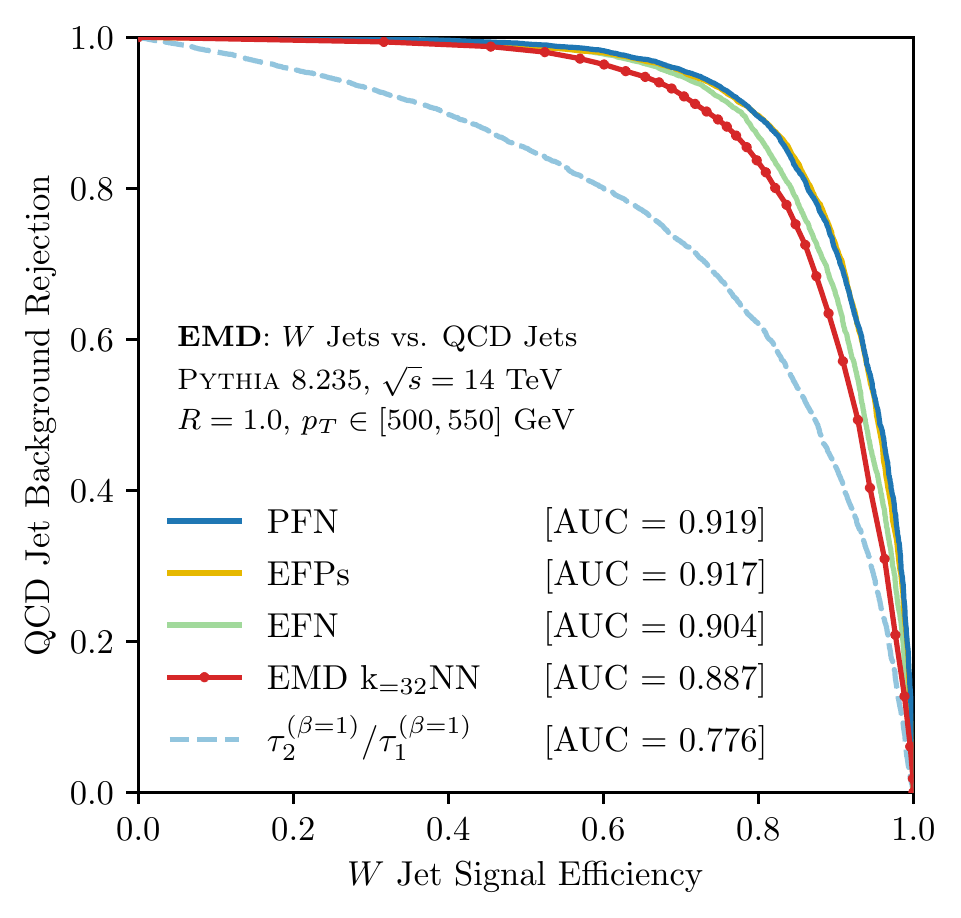}
\caption{ROC curves showing the boosted $W$ classification performance of a $k=32$ nearest-neighbor EMD classifier, which requires no choice of observables or parametrized machine learning architectures.
The EMD classifier is competitive with machine learning techniques known to be good multi-prong classifiers, such as PFNs, EFNs, and EFPs.}
\label{fig:rocs}
\end{figure}

\begin{figure}[t]
\centering
\includegraphics[width=0.98\columnwidth]{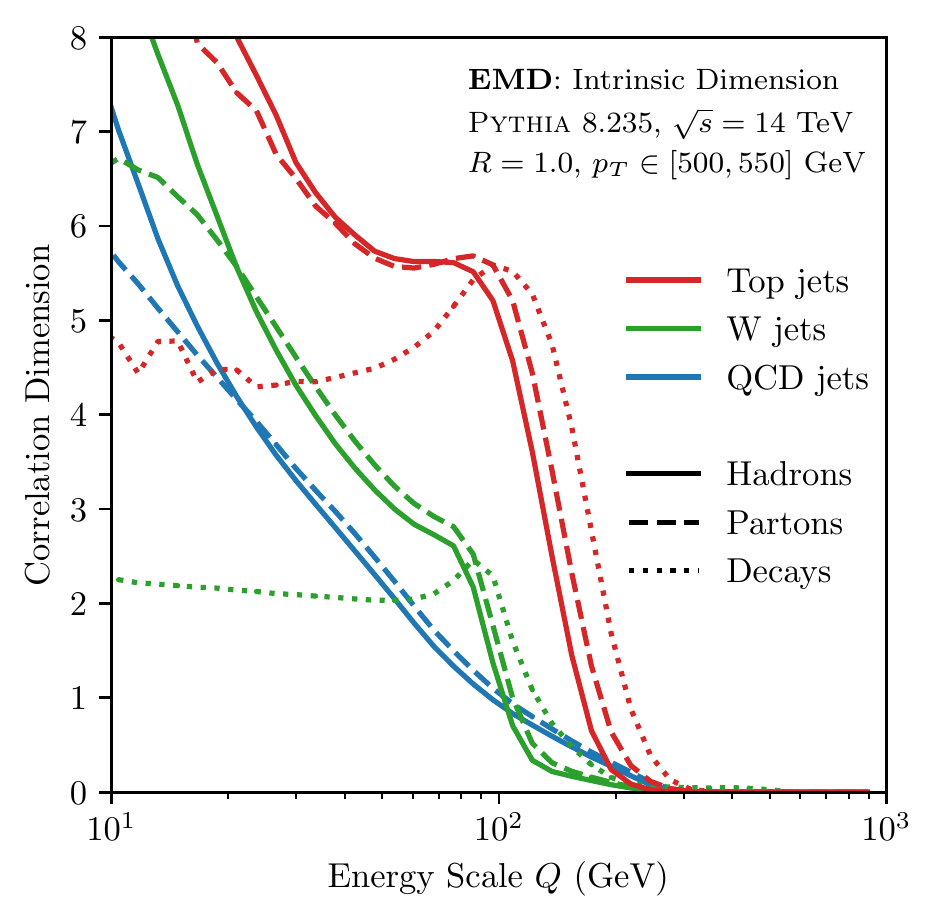}
\caption{The correlation dimension of top, $W$, and QCD jets as a function of the energy scale $Q$ using hadrons (solid), partons (dashed), and hard decay products (dotted).
Generally, QCD jets are the lowest dimensional and top jets are the highest dimensional.
By comparing partons and hadrons, one sees that hadronization affects the structure of the space at scales below about 30 GeV.
Similarly, the hard decay structure of top and $W$ jets governs their dimension at high scales.
Below about 10 GeV, the data become very high dimensional and sparse, making dimension estimation difficult.}
\label{fig:intdim}
\end{figure}

\begin{figure}[t]
\centering
\includegraphics[width=\columnwidth]{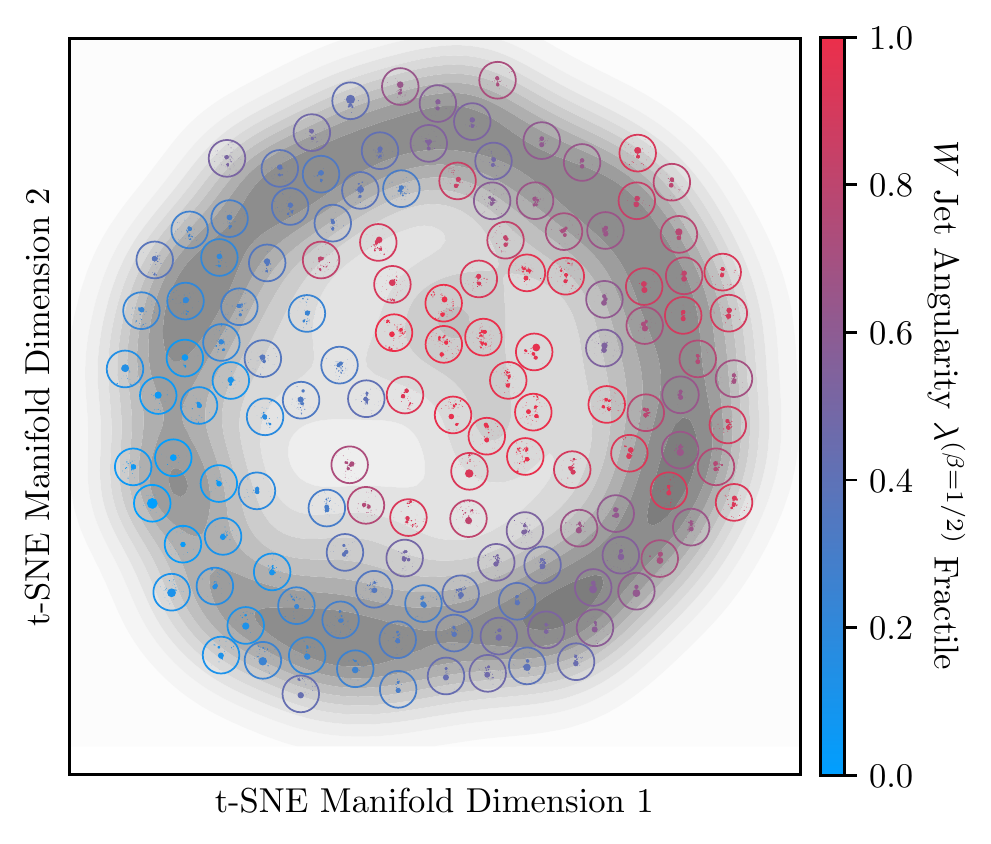}
\caption{The embedding of $W$ jets into a two-dimensional space with t-SNE.
The gray contours represent the density of embedded jets.
Examples of $W$ jets are shown throughout the space.
The color of each jet corresponds to its angularity $\lambda^{(\beta=1/2)}$ fractile to quantify the energy sharing of the two prongs.
An annulus emerges with jets in the lower (upper) region of the manifold having a more energetic lower (upper) subjet.
More complex topologies with the largest angularity values populate the center of the manifold.}
\label{fig:tsne} 
\end{figure}

A metric space is also useful for classification without requiring specially designed observables or parametrized machine learning algorithms.
One of the simplest examples of a non-parametric classifier is the $k$-nearest neighbor (kNN) algorithm~\cite{DBLP:journals/tit/CoverH67}, whereby a given event's closest $k$ neighbors in a reference set are used to determine class membership.
We build a kNN classifier applied to the problem of discriminating $W$ jets from QCD jets using a balanced training sample of 100k total jets.
The classifier output is the number of $W$ jets among the $k=32$ nearest neighbors by EMD.
This method should approach the optimal IRC-safe classifier with a sufficiently large dataset.
The performance of the resulting EMD kNN classifier is shown in \Fig{fig:rocs} as a receiver operating characteristic (ROC) curve, with the Area Under the ROC Curve (AUC) also shown.
For comparison, we include an Energy Flow Network (EFN) and a Particle Flow Network (PFN)~\cite{Komiske:2018cqr} as well as a linear classifier trained on Energy Flow Polynomials (EFPs)~\cite{Komiske:2017aww}.
All classifiers are trained on a 100k training sample and evaluated on a 20k test sample, with the neural networks using 20\% of the training sample for validation and a batch size of 125 (see \Ref{Komiske:2018cqr} for additional details).
The kNN approaches the performance of these state-of-the-art classifiers and significantly outperforms a ratio $\tau_2^{(\beta=1)}/\tau_1^{(\beta=1)}$ of $N$-subjettiness observables~\cite{Thaler:2010tr,Thaler:2011gf} designed to identify two-prong substructure.
It is expected that the performance of the kNN method would improve with more sophisticated kernel density estimation techniques.

It is worth noting that while searching through a large reference set of events to find neighbors naively requires every possible pairwise comparison, in a metric space the triangle inequality can provide a great deal of simplification.
Specialized data structures known as metric trees~\cite{DBLP:journals/ipl/Uhlmann91,DBLP:conf/soda/Yianilos93,DBLP:conf/vldb/Brin95,DBLP:journals/tods/BozkayaO99} have been developed to achieve query times that are approximately logarithmic in the size of the dataset.
While we use direct searches throughout this letter, this is not a fundamental limitation and we leave metric tree query optimizations to future work.

Once a space has been equipped with a metric, it is natural to ask about the structure of the induced manifold.
The most basic aspect of the manifold underlying the data is its dimension, and several notions of its intrinsic dimension exist~\cite{CAMASTRA20032945}.
The correlation dimension~\cite{Grassberger:1983zz,DBLP:conf/nips/Kegl02}, a type of fractal dimension, is suitable for our purposes and is defined using only pairwise distances:
\begin{equation}\label{eq:corrdim}
\dim(Q) = Q\frac{\partial}{\partial Q}\ln\hspace{-0.5em}\sum_{1\le k < \ell \le N}\hspace{-0.5em}\Theta(\text{EMD}(\mathcal E_k, \mathcal E_\ell) < Q),
\end{equation}
where $N$ is the total number of events and the summand indicates whether event $k$ is within EMD $Q$ of event $\ell$.

The correlation dimension is an intrinsically scale-dependent quantity, which is particularly useful as we anticipate different physical effects to dominate jets at different scales.
Shown in \Fig{fig:intdim} is the intrinsic dimension of our top, $W$, and QCD samples over energy scales $Q$ ranging from 10 GeV to 1000 GeV obtained from \Eq{eq:corrdim} with 25k jets.
At high energy scales $Q$, the EMD is governed by the hard decay kinematics, resulting in a relatively simple manifold with low intrinsic dimension.
At energy scales $Q$ approaching the fragmentation and hadronization scales, the structure of the events becomes increasingly complex and the dimension correspondingly increases.
It is satisfying that the dimension is relatively low for a wide range of relevant energies, which is critical for a variety of metric-based techniques such as classification and low-dimensional visualization to work effectively with a realistic amount of data.

Beyond probing its dimension, the entire space of jets can be visualized using techniques such as t-Distributed Stochastic Neighbor Embedding (t-SNE)~\cite{vanDerMaaten2008,DBLP:journals/jmlr/Maaten09,
DBLP:journals/ml/MaatenH12,DBLP:journals/jmlr/Maaten14}, which finds a low-dimensional embedding of the data that attempts to respect the distances between points.
\Fig{fig:tsne} shows a t-SNE embedding of 5k $W$ jets with $p_T\in [500,510]$ GeV into a two-dimensional manifold using {\tt scikit-learn}~\cite{scikit-learn}.
The narrower $p_T$ range focuses the EMD on the jet substructure and was found to yield sharper visualizations, with other choices also yielding sensible results.
The $W$ jets populate a circular subspace roughly corresponding to the energy sharing of the two prongs.
As the $W$ jet originates from a resonant decay, the two decay quarks (after rotation) are solely described by their energy sharing, which satisfyingly emerges from the manifold of $W$ jets.
Moreover, the center of the ring, distant from the annulus, tends to contain the most complex jet topologies, resulting in a type of automatic anomaly detection.

\begin{figure}[t]
\centering
\includegraphics[width=0.88\columnwidth]{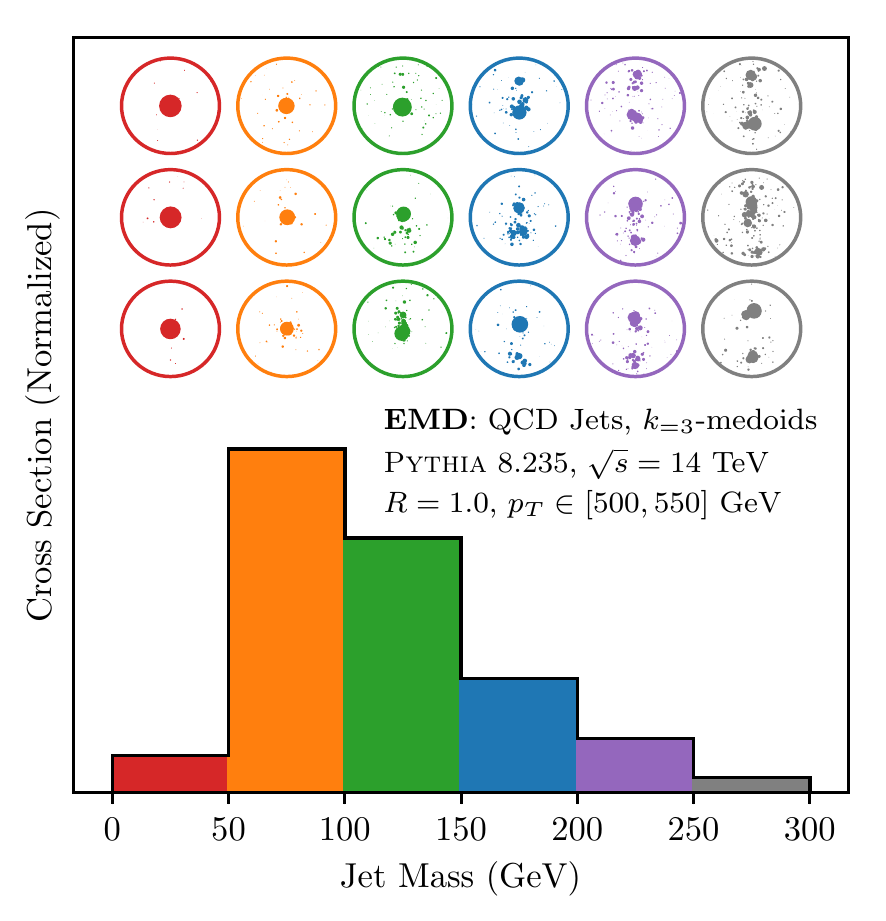}
\caption{The jet mass distribution for QCD jets, with $k=3$ medoids shown above each bin.
This visualization highlights that simple one-prong topologies dominate low jet masses and complex two-prong topologies exist at high jet masses.}
\label{fig:medoids}
\end{figure}

Finally, we illustrate the use of EMD for a new kind of visualization strategy that clusters events to better understand observable distributions.
To describe a given set of events, such as those in a histogram bin, we find the $k$ events (called medoids) which best describe the set in that the sum of distances of each event to its closest medoid is minimized.
This procedure works for any observable and provides an immediate glimpse of the types of event topologies that correspond to a given observable value.
We use an iterative approximation of $k$-medoids from the {\tt pyclustering} Python package~\cite{andrei_novikov_2018_1491324}.
As an illustration, \Fig{fig:medoids} shows the jet mass for QCD jets with $k=3$ medoids per bin, providing a snapshot of the different event topologies at different masses.

In conclusion, we have equipped the space of events with a metric, thereby allowing a powerful suite of new tools and techniques to be directly applied to collider physics.
There are many potential applications of the EMD at colliders beyond those presented here.
Pileup mitigation or detector reconstruction could use the EMD to benchmark performance and thus benefit from the quantitative bounds on IRC-safe observable modifications.
Further, machine learning models could be trained to optimize the EMD, related to recent efforts in generative modeling~\cite{DBLP:journals/corr/ArjovskyCB17,Erdmann:2018kuh,Erdmann:2018jxd,Chekalina:2018hxi}.
By counting neighbors, one could also perform density estimation in the space of events~\cite{Andreassen:2018apy}.
While we have focused on jet substructure, analogous studies could be carried out at the event level, which may require working with composite objects such as jets for realistic computation times.
It would be interesting to explore an EMD strategy for unfolding by matching detector-level and simulated events.
One might consider alternatives to the EMD, such as symmetry-projected metrics~\cite{DBLP:conf/gsi/PeleT13} or $p$-Wasserstein metrics~\cite{wasserstein1969markov,dobrushin1970prescribing} beyond our $p=1$ case, though our conclusions should hold for any physically sensible metric.
Further, using the EMD for model-independent anomaly detection~\cite{Collins:2018epr,DeSimone:2018efk,Hajer:2018kqm,Heimel:2018mkt,Farina:2018fyg,Cerri:2018anq,Collins:2019jip} by finding isolated or clustered event topologies could empower searches for physics beyond the Standard Model at the LHC.

\begin{acknowledgments}

We would like to thank Felice Frankel, Marat Freytsis, Paul Ginsparg, Aram Harrow, Gregor Kasieczka, Andrew Larkoski, Katherine Liu, Benjamin Nachman, Miruna Oprescu, Katherine Quinn, and Jonathan Walsh for helpful discussions.
We benefited from the hospitality of the Harvard Center for the Fundamental Laws of Nature, the Fermilab Distinguished Scholars program, and the Aspen Center for Physics.
This work was supported by the Office of Nuclear Physics of the U.S. Department of Energy (DOE) under Grant No. DE-SC0011090 and the DOE Office of High Energy Physics under Grant Nos. DE-SC0012567 and DE-SC0019128.
JT is supported by the Simons Foundation through a Simons Fellowship in Theoretical Physics.
Cloud computing resources were provided through a Microsoft Azure for Research award and through a Google Cloud allotment from the MIT Quest for Intelligence.
Optimal transport provided by the 2018 Nissan Think Tank.

\end{acknowledgments}

\bibliography{emd}

\end{document}